\documentclass[twocolumn,tighten]{aastex61}
\usepackage{amssymb, amsmath,framed}
\usepackage{tikz}
\usepackage{subfigure}

\usepackage{bm}
\expandafter\ifx\csname package@font\endcsname\relax\else
 \expandafter\expandafter
 \expandafter\usepackage
 \expandafter\expandafter
 \expandafter{\csname package@font\endcsname}
\fi
\def\bq{\begin{equation}}
\def\eq{\end{equation}}
\def\bqy{\begin{eqnarray}}
\def\eqy{\end{eqnarray}}

\begin{document}
\title{Impact and mitigation strategy for future solar flares}

\correspondingauthor{Manasvi Lingam}
\email{manasvi.lingam@cfa.harvard.edu}

\author{Manasvi Lingam}
\affiliation{Harvard-Smithsonian Center for Astrophysics, 60 Garden St, Cambridge, MA 02138, USA}
\affiliation{John A. Paulson School of Engineering and Applied Sciences, Harvard University, 29 Oxford St, Cambridge, MA 02138, USA}
\author{Abraham Loeb}
\affiliation{Harvard-Smithsonian Center for Astrophysics, 60 Garden St, Cambridge, MA 02138, USA}

\begin{abstract}
It is widely established that extreme space weather events associated with solar flares are capable of causing widespread technological damage. We develop a simple mathematical model to assess the economic losses arising from these phenomena over time. We demonstrate that the economic damage is characterized by an initial period of power-law growth, followed by exponential amplification and eventual saturation. We outline a mitigation strategy to protect our planet by setting up a magnetic shield to deflect charged particles at the Lagrange point L$_1$, and demonstrate that this approach appears to be realizable in terms of its basic physical parameters. We conclude our analysis by arguing that shielding strategies adopted by advanced civilizations will lead to technosignatures that are detectable by upcoming missions. 
\end{abstract}

\section{Introduction}
Ever since the famous Carrington flare \citep{Carr59} over $150$ years ago, much attention has been devoted to understanding flares and associated extreme space weather events from a theoretical \citep{Pri14,CLHB} and observational \citep{WH12,Benz17} standpoint. The past two centuries have also been characterized by remarkable advancements in the realms of science and technology.  

Over the past few decades, it has become increasingly apparent that greater reliance on technology has also made us more vulnerable to risks posed by extreme space weather events \citep{Hap11,East17}. For instance, coronal mass ejections can produce powerful geomagnetic storms that disrupt a wide range of electrical systems \citep{Pul07}. In particular, if the Carrington event were to occur now, it would wreak significant damage to electrical power grids, global supply chains and satellite communications. The cumulative worldwide economic losses could reach up to $\$10$ trillion dollars \citep{SSB09,Sch14}, and a full recovery is expected to take several years. 

When the above predictions are viewed in conjunction with the prospect that a Carrington-like storm has a $\sim 10\%$ chance of occurring within the next decade \citep{Ril12}, it is manifestly evident that the risks posed by large flares, and even solar superflares \citep{Mae12,KS13}, should be taken seriously. In this Letter, we present an economic model that assesses the economic losses due to solar flares and propose a strategy by which the damage could be mitigated.

\section{The economic impact of extreme space weather events}
We present a heuristic model that quantifies the potential economic damage caused by extreme space weather events associated with large flares (or superflares) in the future and explore the implications. 

\subsection{A mathematical model for economic losses}
The analysis of \emph{Kepler} data from $10^5$ stars enabled the determination of the occurrence rate of superflares $N$ as a function of their energy $E$ \citep{Mae12}. For G-type stars like the Sun, it was shown that
\begin{equation} \label{Freq}
    \frac{d N}{d E} \propto E^{-\alpha} \quad \alpha \sim 2.
\end{equation}
From the above relation, it is evident that the typical wait time $\tau$ for a superflare with energy $E$ to occur satisfies $\tau \propto E^{\alpha - 1}$. Since superflares have been predicted to occur on the Sun over fairly short timescales \citep{Uso13,KS13}, (\ref{Freq}) may also be applicable to our star. A similar power-law scaling relation, albeit with a slightly different value of $\alpha$, exists for solar flares \citep{Shim95,AP02}. 

Flares are associated in many instances with coronal mass ejections (CMEs) and stellar energetic particles (SEPs), both of which are capable of causing significant economic damage \citep{East17}. By denoting the total economic damage (in USD) by $D$ and following the prescription in \citet{Ling2017}, we adopt $D \propto E$, which leads to a wait time scaling
\begin{equation} \label{Dam}
    \tau \propto D^{\alpha - 1}
\end{equation}
We denote the Gross Domestic Product (GDP) at the present time by GDP$_0$ and introducing the dimensionless variable $x = D/\mathrm{GDP}_0$.\footnote{We will henceforth work with the parameters for the USA because some of the corresponding data for the entire world is not readily available ($\mathrm{GDP}_0 \sim 20$ trillion USD).} We rewrite (\ref{Dam}) as a differential equation in terms of normalized variables,
\begin{equation} \label{DiffDam}
    \frac{dx}{d \tau} = \frac{x^{2-\alpha}}{\tau_s \left(\alpha - 1\right)},
\end{equation}
where $\tau_s$ is related to the constant of proportionality in (\ref{Freq}); all timescales will be measured in units of years and are therefore dimensionless. In order to estimate $\tau_s$, we note that a superflare with energy $\sim 10^{34}$ ergs has been predicted to occur on the Sun once every $\sim 2000$ years \citep{Shi13}. In contrast, the Carrington 1859 flare had an energy $\sim 5 \times 10^{32}$ erg \citep{CD13} and would result in economic losses that are $\sim 10\%$ of GDP$_0$ if it occurred now \citep{SSB09}. Choosing $\alpha \sim 2$ and using (\ref{Dam}) along with the above data, we end up with $\tau_s \sim 1000$.

Relations (\ref{Dam}) and (\ref{DiffDam}) were based on the premise that the GDP and the economic damage are purely \emph{static}, i.e. larger economic losses result only because the magnitude of the expected flare (and concomitant economic damage) increases over longer time intervals. In reality, it is important to recognize that the economic losses will also amplify due to another factor - technological advances render our systems more fragile to space weather events \citep{Bot06}. This explains why the Carrington event will lead to significantly more damage currently than it did at its actual time of occurrence.

Thus, we must account for the rise in technological sophistication by including another term in (\ref{DiffDam}). If technology were to obey the law of exponential growth, one should include a term $x/\tau_p$, where $\tau_p$ is the $e$-folding timescale associated with technological progress. However, exponential growth \emph{ad infinitum} is unrealistic, and we shall instead suppose that the growth (and concomitant damage) is logistic in nature based on widely predicted trends \citep{Mont78,GNN99} and the dynamical behavior of human capital \citep{SW97}. When written as an ODE \citep{AbSte65}, logistic growth obeys
\begin{equation} \label{DiffLog}
     \frac{dx}{d \tau} = \frac{x}{\tau_p}\left(1 - \frac{x}{\Gamma_m}\right),
\end{equation}
where $\Gamma_m$ is the value at which $x$ would saturate. Upon combining (\ref{DiffDam}) and (\ref{DiffLog}), the final differential equation for the economic losses is
\begin{equation} \label{FinDiff}
    \frac{dx}{d\tau} = \frac{x^{2-\alpha}}{\tau_s \left(\alpha - 1\right)} + \frac{x}{\tau_p}\left(1 - \frac{x}{\Gamma_m}\right),
\end{equation}
In order to determine the value of $\tau_p$, we note that economic losses arising from geomagnetic storms (associated with flares) are primarily due to damage caused by electrical power grids \citep{Hap11}. Since worldwide electricity generation in the near-future is expected to increase by $1.9\%$,\footnote{\url{https://www.eia.gov/outlooks/ieo/electricity.php}} we hypothesize that the growth of electrical grids is approximately equal to the increase in electrical energy generation. This assumption amounts to choosing $\tau_p \sim 50$. The only unknown parameter at this stage is $\Gamma_m$. The current year corresponds to $\tau=0$, and we specify the initial condition $x(0)=0$. 

\begin{figure*}
$$
\begin{array}{cc}
  \includegraphics[width=7.5cm]{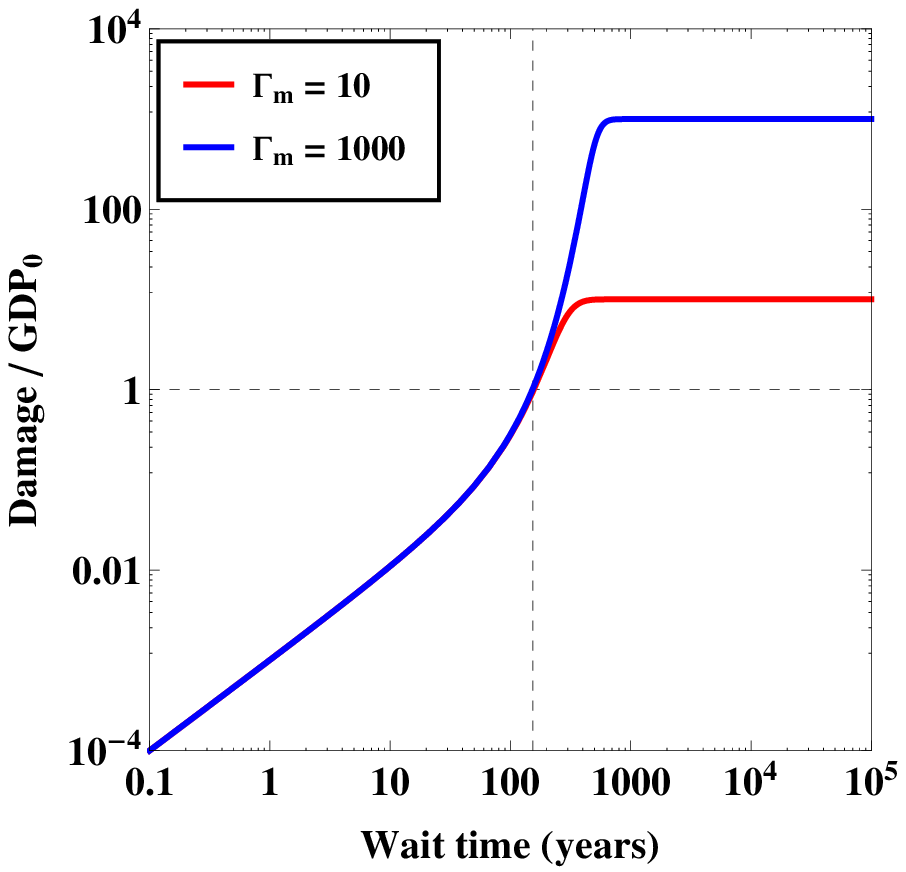} &  \includegraphics[width=7.1cm]{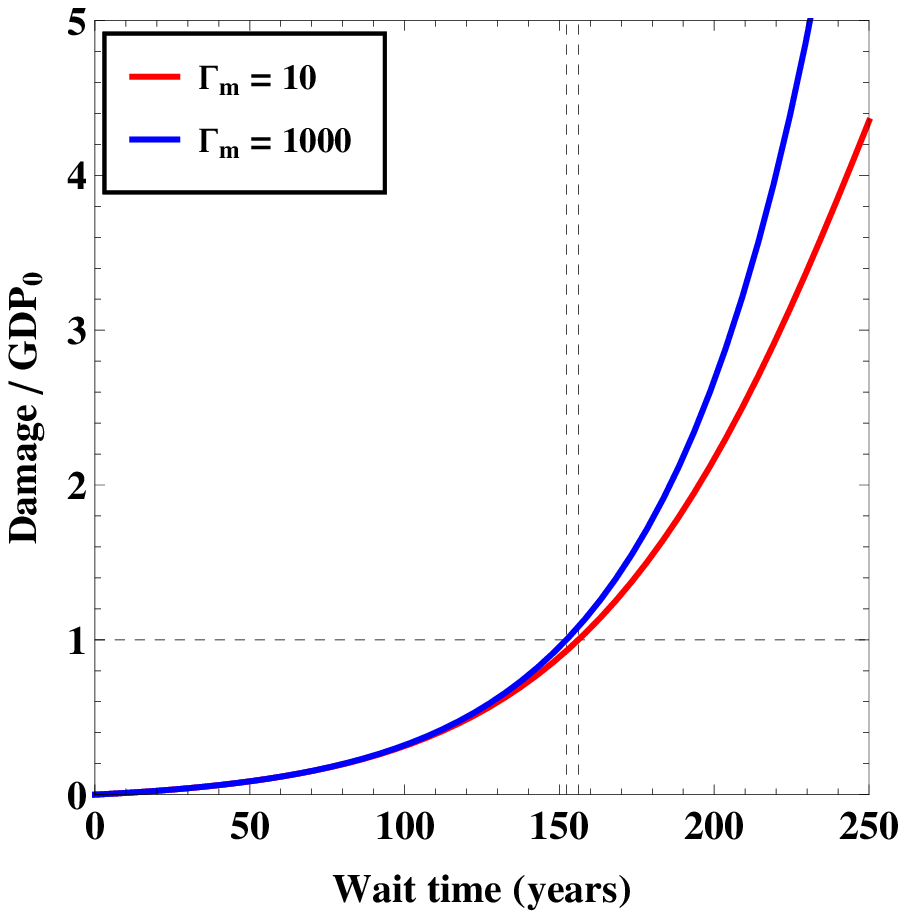}\\
\end{array}
$$
\caption{The economic losses (normalized by current GDP of the USA) as a function of the wait time (in units of years). The dotted black lines signify the wait times at which the economic damage caused by the flare equals that of the current GDP. The left-hand panel depicts the log-log plot and the global picture, while the right-hand panel shows a linear scale on its axes with the power-law and exponential growth stages.}
\label{FigFlare}
\end{figure*}

For the above choices of the free parameters, the differential equation can be solved analytically by recognizing that (\ref{FinDiff}) corresponds to a particular case of the Riccati equation \citep{BenOr78}. Hence, it can be converted into a \emph{linear} second-order differential equation with constant coefficients which can be solved in a straightforward manner \citep{AbSte65}. The final solution is:
\begin{equation} \label{AnalSol}
    x(\tau) = \frac{\exp\left(q \tau/50\right)-1}{10\left[q + 1 + \left(q-1\right)\exp\left(q \tau/50\right)\right]},
\end{equation}
where $q^2 = 1 + 1/\left(5\Gamma_m\right)$. It is clear that $x(0)=0$ and $x(\infty) = \Gamma_m \left(q + 1\right)/2$. For $\Gamma_m > 1$, we see that $x(\infty) \approx \Gamma_m$, thereby indicating that the asymptotic limit is dominated by the ``cap'' imposed by the logistic growth of technology, which corresponds to the second term in the RHS of (\ref{FinDiff}). The solution has been graphically plotted for the values of $\Gamma_m = 10$ and $\Gamma_m = 1000$ in Fig. \ref{FigFlare}. One can relax the assumption that $\Gamma_m$ is a constant, and solve (\ref{FinDiff}) numerically by choosing some ansatz $\Gamma_m \equiv \Gamma_m(\tau)$. However, the technological growth (and damage) would no longer obey the well-established paradigm of logistic growth, as discussed earlier, and we not consider such cases herein.

One can carry out a similar analysis by extrapolating to the past. There are two competing effects that must be taken into account. Going further back in time enables a larger flare to occur that would result in greater economic losses. However, the potential for increased damage is counteracted by the fact that the level of technology was lower. In fact, if one were to proceed beyond 200 years into the past, large flares would have had a minimal economic impact because there were no electrical grids, satellites and global supply chains \citep{Bai07}. One can therefore estimate the maximum economic damage that could have been caused in the fact by determining the point $\left(x_\star,\tau_\star\right)$ where the two competing effects in (\ref{FinDiff}) balance each other. This amounts to solving ${1}/{1000} \approx {x_\star}/{50}$. Thus, the maximum economic damage possible would have been about $5\%$ of the current GDP and occurred $50$ years ago. Moving further back in time, the economic losses would be dominated by \emph{exponential damping} (due to technological regression), thus explaining why the damage due to the 1859 Carrington flare was minimal.

\subsection{The implications for the future}
From Fig. \ref{FigFlare} and (\ref{AnalSol}), a wide range of inferences can be drawn, and the implications are described below.
\begin{enumerate}
    \item Regardless of the choice of $\Gamma_m$, we find that the initial behavior ($\lesssim 50$ years) is governed by the first term in the RHS of (\ref{FinDiff}). This observation can be explained by noting that, in this period, the growth in technology is not particularly significant. Instead, most of the risk stems from the fact that a longer wait time leads to a bigger flare, and more economic damage.
    \item The point at which $x = 1$ occurs requires a wait time of $150$ years. It is virtually independent of the value chosen for $\Gamma_m$ as seen clearly from the right-hand panel of Fig. \ref{FigFlare}.
    \item After the power-law stage, the second term in the RHS of (\ref{FinDiff}) becomes more important. Hence, in this interval, most of the economic damage arises from advancements in technology. The losses resulting from this factor, which grow exponentially until saturation, dominate over the damage that would result from just the first term since bigger flares occur over increasingly sporadic intervals.
    \item However, once the technological growth slows down and saturates, the economic losses also undergo saturation. Thus, even though more energetic flares are expected for longer wait time intervals, the damage achieves a ``bottleneck'' that is imposed by technological constraints. 
    \item The onset of saturation is weakly dependent on the choice of $\Gamma_m$ as seen from the left-hand panel of Fig. \ref{FigFlare}; it depends primarily on the value of $\tau_p$.
    \item The vulnerable phase is during the relatively short-lived regime of exponential amplification that is likely to begin a few decades henceforth. Hence, the ideal scenario entails the identification and implementation of an effective strategy to mitigate the risks from extreme space weather events within the next century. 
\end{enumerate}
Before proceeding further, we note that Fig. \ref{FigFlare} and (\ref{AnalSol}) should be perceived as \emph{upper} bounds for the resultant damage, since our work implicitly assumes that flares (and superflares) are accompanied by CMEs and SEPs that directly impact the Earth and cause economic losses; in reality, these phenomena are not necessarily correlated with one another. 

\section{Mitigation strategies for extreme space weather events} \label{SecMit}
From the preceding discussion, it is evident that large flares will cause significant economic damage (close to the current GDP) even within the next century. Although this fact has been thoroughly documented \citep{Hap11,East17}, the risks from flares and superflares have not received the same level of attention as asteroid impacts and Earth-based natural catastrophes \citep{Pos04}.

Hence, we explore a few potential strategies that may be employed in the future to mitigate the risks posed by large flares. These approaches have also been investigated in the context of spacecraft shielding, i.e. to protect astronauts from charged particles during long-term missions \citep{Spil07}. The first option is to enforce shielding by using an appropriate material (e.g. water), but the weight would be extremely prohibitive. The second alternative is based on electrostatic shielding, but it has been argued to be ineffective as it would strongly attract electrons (while repelling cations) and entails significant power requirements \citep{Park06}.

\begin{figure*}[!ht]
\centering
\includegraphics[scale=0.5]{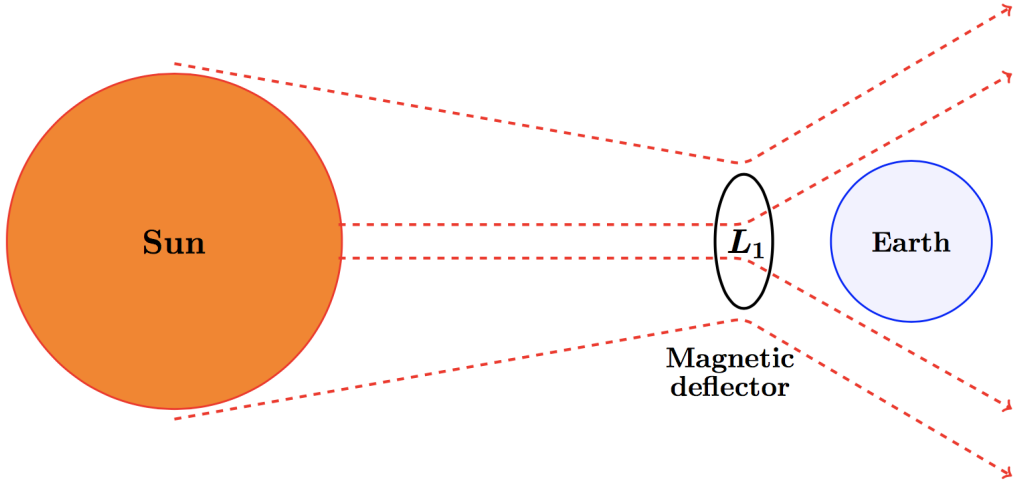}
\caption{A schematic figure of the proposed magnetic shield that will be placed at the Lagrange point $L_1$ to deflect SEPs emitted by the sun during extreme space weather events (not drawn to scale).}
\label{FigSchem}
\end{figure*}

This leads us to magnetic shielding wherein particles are deflected by means of the Lorentz force. Although several reasons exist as to why this option is not feasible for space missions \citep{ShKr07}, most stem from the fact that the spacecraft is compact and that the magnetic fields required for deflection would be extremely high. In contrast, we will consider a scenario wherein the ``magnetic deflector'' is placed at a certain distance away from the Earth. The deflection required would be comparatively small, as would the corresponding magnetic fields.

The schematic diagram for this proposal is presented in Fig. \ref{FigSchem}. The idea is to place the deflector at the Lagrange point L$_1$, which lies between the Sun and the Earth. The radius of gyration $r_L$ is
\begin{equation} \label{GyrDef}
    r_L \approx \frac{E_p}{eB},
\end{equation}
where $B$ is the magnetic field, and $E_p = \gamma m_p c^2$ is the relativistic energy of a typical SEP. The gyration radius $r_L$ should be calibrated such that it is comparable to, or smaller than, the distance between L$_1$ and Earth denoted by $d = 1.5 \times 10^{11}$ cm. Using $r_L \lesssim d$ in conjunction with (\ref{GyrDef}) yields
\begin{equation} \label{Bfield}
    B \gtrsim 2.2 \times 10^{-5}\, \mathrm{G}\,\left(\frac{E_p}{1\,\mathrm{GeV}}\right).
\end{equation}
If we suppose that the radius $R$ of the loop is comparable to the radius of the Earth $R_\oplus$, and make use of the Biot-Savart law, we find
\begin{equation}
    B \propto \frac{I}{R},
\end{equation}
and we can solve for the current $I$ using (\ref{Bfield}). We arrive at $I \sim 2.2 \times 10^4$ A; note that $I$ refers to the \emph{total} current, and the current per coil turn can be much lower depending on the number of turns. Using the scaling for the magnetic moment $\mu$, namely $\mu \propto I R^2$, we conclude that the total magnetic moment of the deflector system would be approximately $10^{-4}$ that of the Earth.

The power dissipated in the wire can be computed for a given material through Ohm's law
\begin{equation} \label{Pow}
    P = I^2 \mathcal{R},
\end{equation}
where $\mathcal{R}$ is the resistance and is given by
\begin{equation} \label{Resis}
    \mathcal{R} = \frac{\rho \mathcal{L}}{\mathcal{A}},
\end{equation}
where $\mathcal{L} = 2\pi R$ and $\mathcal{A} = \pi r^2$ are the length and cross-sectional area of the wire, respectively. Let us suppose that the wire were made of copper and has a radius  $r = 1$ cm. We end up with $\mathcal{R} \approx 2.1\, \mathrm{k}\Omega$ by using (\ref{Resis}). Substituting this value of $R$ into (\ref{Pow}) yields $P \approx 1$ TW, which is about $10\%$ of the world's current power consumption. If we suppose that the power is radiated away as per the black body prescription, we end up with
\begin{equation}
    P = 2 \pi \sigma\, r\mathcal{L}\, T^4,
\end{equation}
where $T$ is the effective black body temperature. We obtain $T = 1627$ K for this system, somewhat higher than the melting point of copper, implying that $r = 1$ cm is an upper bound; a thinner wire would give rise to a value of $T \propto r^{-3/4}$ higher than the melting point of metals. For our choice of parameters, the mass of the copper coil would be $\sim 10^5$ tons. If a superconducting current loop were employed instead, the power dissipated would be lower, even when the thickness (and mass) of the coil is significantly reduced.

The total cost involved in lifting a $10^5$ ton object into space would be around $\$ 100$ billion, assuming that the payload cost per kg is $\$ 1000$. This value is comparable to the total cost of the International Space Station, and is $3$-$4$ orders of magnitude lower than the current world GDP, or the economic damage from a flare $\sim 100$ years henceforth. Alternatively, one could mine the asteroid belt and construct the superstructure directly in space.\footnote{Exocivilizations implementing this activity might produce detectable technosignatures \citep{FE11}.}

Let us suppose that the incident solar energetic proton flux is $\Phi$ and the deflection angle is $\theta$. Hence, we would expect the total force exerted on the wire to be
\begin{equation}
    F = \Phi \left(\gamma m_p c \sin \theta\right) \left(\pi R^2 \right),
\end{equation}
where the second factor constitutes the change in momentum per particle, and the last factor represents the area enclosed by the loop. We compute the pressure on the wire $\mathcal{P} = F/\left(2\pi r \mathcal{L}\right)$, and estimate the strain:
\begin{equation}
    \frac{\Delta \ell}{\ell} = \frac{\mathcal{P}}{Y},
\end{equation}
where $Y$ is the Young's modulus of the wire material (e.g. copper). We choose a characteristic value of $\Phi$ equal to that of the Carrington flare \citep{CD13}. Even for maximum deflection, we find $\Delta \ell/\ell \sim 8 \times 10^{-17}$, indicating that the resultant material strain is negligible. If we replace $\mathcal{P}$ with the ram pressure, which models the bulk non-relativistic outflow, during the Carrington event \citep{NPKG}, the strain remains insignificant since we obtain $\Delta \ell/\ell \sim 3 \times 10^{-12}$.

Our proposal is not meant to be exhaustive from an engineering standpoint. For instance, most magnetic shielding devices are not effective at deflecting particles along the axis \citep{Park06}; a potential solution is to incorporate orthogonal/inner current loops that partially offset this deficiency. Moreover, an energy source will be required for maintaining the current although, in principle, it could be extracted from the sun by setting up photovoltaic panels in space. 

\section{Implications for SETI} \label{SecSETI}
The \emph{Kepler} mission has established that superflares are more common on M- and K-dwarfs when compared to G-type stars \citep{Mae12}. As such stars constitute the majority in our Galaxy, and given the relative ease of detecting exoplanets around them, they represent natural targets in the context of searching for biosignatures and technosignatures. Hence, it seems reasonable to conclude that technologically advanced civilizations on planets orbiting these stars would be well aware of the economic and biological risks posed by flares and superflares \citep{Ling2017}. In order to mitigate the damage wrought by extreme space weather events, it is very conceivable that they would adopt shielding strategies against such phenomena. 

The form of shielding may entail simple magnetic deflectors along the lines discussed in Sec. \ref{SecMit}, or more complex strategies. One such approach, explored by \citet{CV16} in the context of supernovae and Gamma Ray Bursts, entails shielding the planet through a swarm of planetesimals confined by electromagnetic fields. Another possibility is to deliberately ``enhance'' the planet's natural magnetosphere along the lines of artificial mini-magnetospheres currently being studied on Earth \citep{BKB14}. Lastly, it may be possible to employ St{\"o}rmer shielding \citep{Storm55} on a planetary scale, wherein charged particles can be excluded from a toroidal region. 

Each of the above cases would necessitate large-scale engineering projects that may be detectable by future observations. Hence, we contend that mitigation strategies for stellar flares should fall under the purview of Dysonian SETI \citep{BCD11,WCZ16}. We anticipate that the transit light-curves should be considerably altered from those of natural planets \citep{Arn05}.  More specifically, the above solutions display several similarities with ``starshades'' around exoplanets to offset runaway greenhouse effects \citep{Gai17} as these structures are planet-sized and situated at the L$_1$ point. By drawing upon this analogy, we suggest that mutual occultations of shields and Earth-sized planets (during transits) situated around solar-type stars will produce distinctive signatures. These technosignatures may fall within the capabilities of the James Webb Space Telescope (JWST), and other future space observatories.

\section{Conclusions}
In this Letter, we have proposed a basic mathematical model to describe the economic losses resulting from flares. Our model predicted that initial damage arises due to larger flares being associated with longer wait times. The subsequent evolution is governed by the growth in technological infrastructure, and we conclude that a period of exponential amplification is followed by slowing down and eventual saturation. One of our primary conclusions was that the economic losses caused by a flare $150$ years in the future would equal that of the current GDP of the USA.

Motivated by these considerations, we evaluated different mitigation strategies to protect human technological systems from flares, and suggested that magnetic shielding may constitute a viable means of protection. We proposed that an Earth-sized current loop could be set up at the Lagrange point L$_1$ to deflect the harmful charged particles associated with extreme space weather events. We estimated the magnetic field, electrical current, heat dissipated and material strain for this device, and concluded that their values appear to be feasible. 

Next, we briefly considered the existence of such structures around other planet-star systems, and inferred that they would represent distinctive technosignatures of advanced civilizations. These shields will leave imprints in the transit light-curves that could be detectable by future telescopes such as the JWST. 

In light of mounting evidence indicating that the economic damage due to large flares would be very extensive, we hope that this work will stimulate further research to identify methods for protecting our planet that are both physically sound and technologically implementable within the next century.

\acknowledgments
This work was partly supported by grants from the Breakthrough Prize Foundation for the Starshot Initiative, Harvard University's Faculty of Arts and Sciences, and the Institute for Theory and Computation (ITC) at Harvard University.


\end{document}